\documentclass[a4paper]{jpconf}								

\usepackage{graphicx}	
\usepackage{epstopdf}

\begin{document}

\title{New model of relativistic slowly rotating neutron stars with surface
layer \textit{crust}: application to giant \textit{glitches} of Vela Pulsar}

\author{L. M. Gonz\'alez-Romero and
J. L. Bl\'azquez-Salcedo}

\address{ Depto. F\'{\i}sica Te\'orica II, Facultad de  Ciencias F\' \i sicas,
Universidad Complutense de Madrid, 28040-Madrid, Spain.}

\ead{mgromero@fis.ucm.es, joseluis.blazquez@fis.ucm.es}

\begin{abstract}

Introducing a surface layer of matter on the edge of a neutron star in slow rigid
rotation, we analyze, from an intrinsic point of view, the junction conditions
that must be satisfied between the interior and exterior solutions of the
Einstein equations. In our model the
core-\textit{crust} transition pressure arise
as an essential parameter in the description of a configuration. As an
application of this formalism, we describe giant \textit{glitches} of the Vela
pulsar as a result of variations in the transition pressure, finding that
these small changes are compatible with the 
expected temperature variations of the inner crust during \textit{glitch} time

\end{abstract}

\section{Introduction}

Many pulsars show sudden spin jumps, \textit{glitches}, superimposed to the
gradual spin down due to the continued loss of angular momentum  suffered by
the star.
The study of the properties of the star during the \textit{glitch} time is
essential to understand the structure of the neutron star that models the
pulsar \cite{Lorenz:1993,Link:1999}.  
In Vela
pulsar, giant \textit{glitches} with relative  period variation of the order
of $10^{-6}$ have been observed. Many mechanisms have
been proposed to explain \textit{glitches}.

The models that deal with \textit{glitches} are associated with the layer
structure of neutron stars. Two regions of
the star can be differentiated: the core and the \textit{crust}. It
is thought that the \textit{crust} region has a solid crystalline structure similar to a 
metal \cite{Haensel:2004nu}. Some theories propose  that the \textit{glitch} is
triggered by the rupture of the \textit{crust} as a
consequence of the tensions on the \textit{crust} that try to adequate the
ellipticity of the \textit{crust} to the changing angular velocity of the star. 

Because of the large density gradient close to the transition between the core
and the \textit{crust}, most of the crustal matter resides in the shells in
the inner part of the \textit{crust} \cite{Pethick:1995}. Furthermore, it 
is noteworthy that the dynamical properties of the neutron star depend
strongly on the transition pressure between the star's core and the
\textit{crust} as have been pointed up by \cite{Lattimer:2001} and
\cite{Cheng:2002}.  Because most of the matter of the \textit{crust} is found in the
shells near the transition region between  the core of the star and the inner
\textit{crust}, this is the region where the properties of the \textit{crust} are relevant
\cite{Pethick:1995}. Hence, we will treat the \textit{crust} 
as a surface layer that envelops the star's core. In the next section we will
describe how our model of slowly rotating neutron star with a surface layer
 \textit{crust} is constructed. 

\section{Construction of models of slowly rotating neutron stars with surface
  \textit{crust}.} 

We assume that the neutron star is in permanent rigid rotation. The star
rotates with constant angular velocity $\Omega$ around an axis, so the
resulting space-time is axisymmetric and stationary. We assume that the
rotation of the star is sufficiently slow and, hence, the 
Hartle-Thorne perturbative solution can be used
\cite{Hartle:1967,Hartle:1968}. To second order in the angular velocity of 
the star $\Omega$, appropriate coordinates
can be chosen so that the metric can be written as follows
$ds^{2}=-e^{2\psi (r)}[1+2h(r,\theta)]dt^{2}+e^{2\lambda
  (r)}[1+2m(r,\theta)]dr^{2}+r^{2}[1+2k(r,\theta)]\left[d\theta ^{2}-\sin
  ^{2}\theta(d\phi -\omega(r)dt)^{2}\right]$.

We have to solve the Einstein equations in the interior of the star, where the
matter is described by a perfect fluid tensor with equation of state
$\rho=\rho(p)$ and in the outer region, where
space is empty.  Also, we have to impose appropriate boundary conditions in
the surface of the star. It is very desirable to introduce a new radial
coordinate $R$ adapted to the surface of the star \cite{Hartle:1967,Hartle:1968}.


We assume that the core of the star is surrounded by a thin surface layer
(simplified \textit{crust} model). Then we find that in the surface of the
star $R=a$, the stress-energy tensor is non null and it can be written as
$T_{c}^{\mu \nu }(R)=\rho _{c}(\theta ) u_{c}^{\mu }u_{c}^{\nu }(a)$, i.e. the
surface layer can be described
as a perfect fluid with its own angular
velocity $\Omega_{c}$. The surface energy density can be written up to second
order as $\rho _{c}(\theta )=\varepsilon +\delta \varepsilon(\theta
)$. We must impose the usual junction conditions on the interior and
exterior solutions, but taking into account the introduction of a surface
layer of matter on the border of the star. We use the intrinsic formulation of
these conditions \cite{Gravitation} \cite{pegado_mio}: the first fundamental
form must be continuous across the surface of the star, and the second
fundamental form presents a discontinuity given by the
stress-energy tensor in the surface of the star. In order to use these
junction conditions in the slow rotation approximation, we expand them in
terms of $\Omega$, finding the following results for each one of the metric
functions:  

\textbf{Order zero:} we obtain important conditions for the mass function and for the pressure of transition between core and \textit{crust}
\begin{eqnarray}
M_{ext}=M_{int}+4\pi a^{2}\sqrt{1-\frac{2M_{int}}{a}}\varepsilon -8\pi
^{2}a^{3}\varepsilon ^{2}  
\label{59}\\
\frac{M_{ext}}{a^{2}\sqrt{1-2M_{ext}/a}}-\frac{M_{int}+4\pi
a^{3}p_{int}}{a^{2}\sqrt{1-2M_{int}/a}} = 4\pi \varepsilon
\label{60}
\end{eqnarray}
The equation (\ref{59}) gives the total mass of the
star to zero order ($M_{ext}$) as the addition of three terms: the core mass
($M_{int}\equiv 4 \pi \int_0^a \rho R^2 d R$), the \textit{crust} mass, and a
negative bounding energy term. 
We interpret (\ref{60})  as an
equation giving the surface density of energy $\varepsilon$ in terms of the total
mass of the star, the interior mass, and the core-\textit{crust} transition
pressure $p_{int}$.
This condition is used to obtain the radius of the star.

\textbf{First order: } we obtain continuity of the rate of rotation of the
inertial frames $\omega(a)_{ext}=\omega (a)_{int}\equiv \omega (a)$, and an expression for the total angular momentum:
 \begin{equation}
 J=a^{6}e^{\lambda(a)_{ext}}(e^{-\lambda(a)_{int}}\left[ \partial _{R}\overline{\omega }(a)\right]
 _{int}+16\pi \varepsilon \overline{\omega_{c} })/6
 \label{conJ}
 \end{equation}
where $\overline{\omega_{c} }=\Omega _{c}-\omega(a)$. It can be seen
\cite{Hartle:1967,Hartle:1968} that the first term of expression (\ref{conJ})
has the same sign as $\Omega$ (in our case always positive). However, the
second term depends on the sign of $\Omega_{c}$. Hence, contra-rotating
configuration are possible (the core and the \textit{crust} contra-rotate),
and even,  we could find certain critical configuration in which the total
angular momentum is null.  

\textbf{Second order: } we obtain $\Delta\lbrack r^{\ast }(a,\theta )]=0$
where $r^{\ast }$ is defined as in 
\cite{Hartle:1967,Hartle:1968}, so both the mean radius and the
eccentricity are continuous. Also, conditions for the second order
perturbation of the mass function and surface density ($\delta\varepsilon$)
are obtained.  

Finally the last of the second order conditions fixes the value of the angular
velocity of the \textit{crust} $\Omega _{c}=\omega(a)  \pm (\Omega- \omega(a))$.
We obtain two possible configurations for the same star core and transition
pressure: one with the \textit{crust} co-rotating with the interior fluid and
with identical velocity, and a special configuration with contra-rotating
\textit{crust}. 

\section{Some properties of the model using realistic equations of
  state for the core}

\begin{figure}[h]
 \begin{minipage}{18pc}
 \centering
 \includegraphics[width=0.87\textwidth]{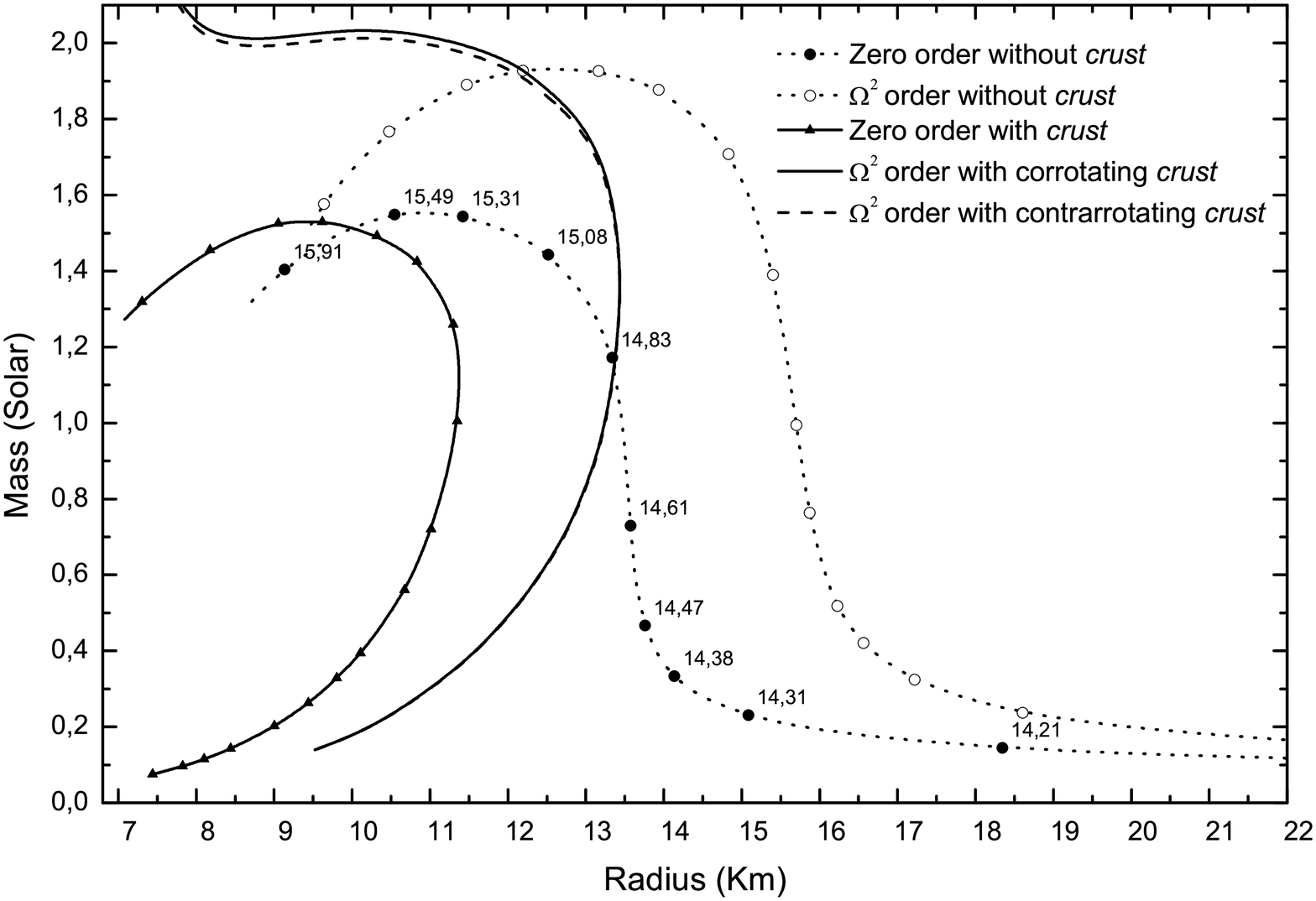}
 \caption{Mass vs radius for static/rotating configurations with a
   10\% \textit{crust} (BPS)}
  \label{MR}
 \end{minipage} 
\begin{minipage}{18pc}
\centering
\includegraphics[width=0.87\textwidth]{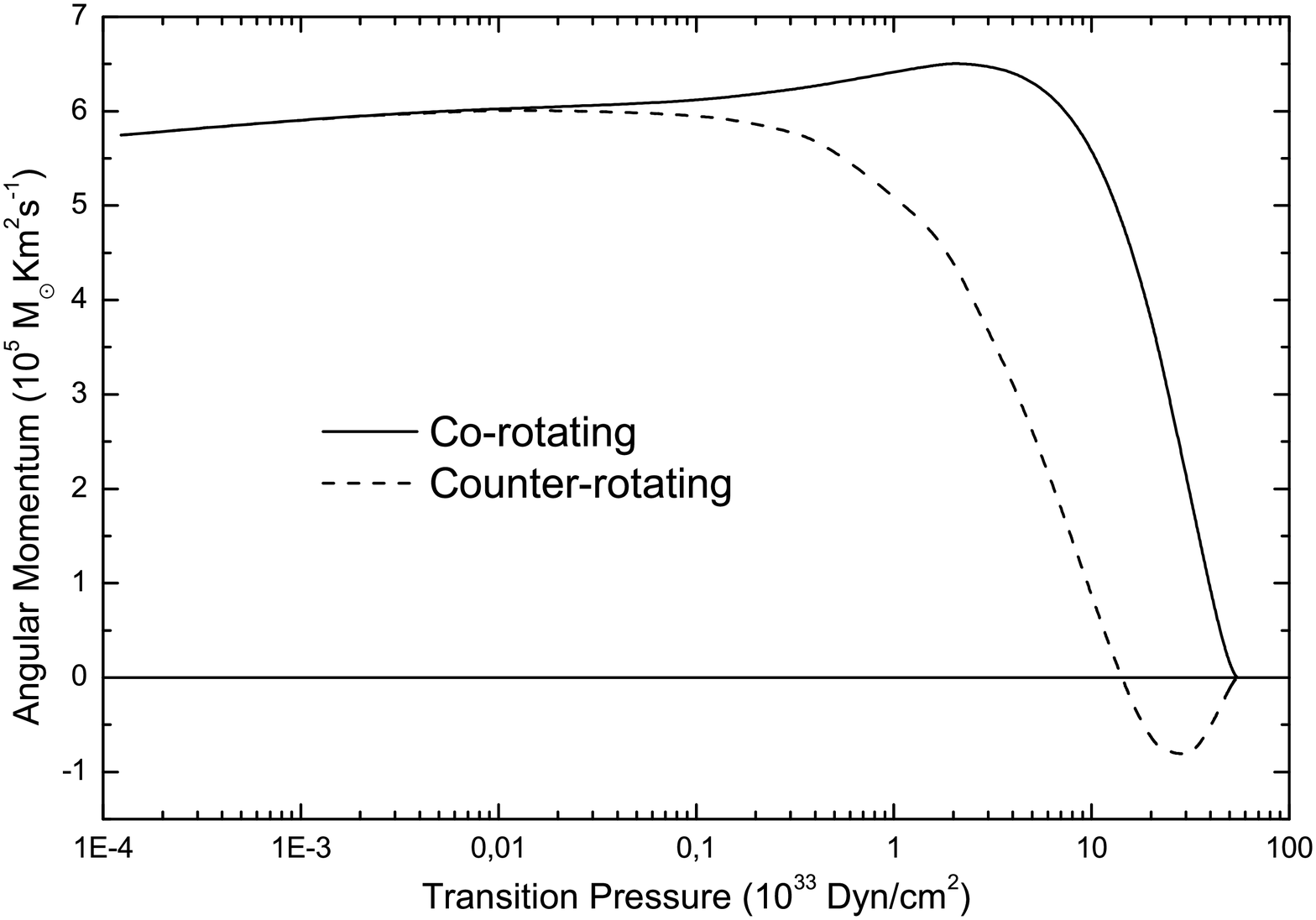}
\caption{Angular momentum vs transition pressure for
  fixed $\rho_{c}$ (BPS).}
\label{Jpint}
\end{minipage}\hspace{3pc}
\end{figure}

To obtain a model of a neutron star with a surface layer
\textit{crust} in permanent slow rotation, for a given equation of state of
the core, we have to fix three parameters. We choose the total mass of the
star, the central density, and the core-\textit{crust} transition
pressure. Then, we integrate the Einstein equations inside the star (using a
realistic equation of state for the core) and we match this solution with
the exterior solution found in \cite{Hartle:1967,Hartle:1968}  using the junction conditions
explained above. We have performed our simulations using an equation of state for high
pressures from \cite[K=240 MeV,$\frac{m^{*}}{m}=0.78$ and
$x_{\sigma}=0.6$]{Glendening:2000}   
and the BPS equation of state for
the lower pressures. Similar results have been obtained for the analytical fit
of the Sly equation presented in \cite{Haensel:2004nu}. 

We present some of our results in figures \ref{MR} and \ref{Jpint}. In fig. \ref{MR}, we show multiple configurations for both
static stars and rotating stars, 
with a \textit{crust} of approximately 10$\%$
of the total mass and mass shedding angular velocity. On the static
configurations without \textit{crust} we write 
the value $\log10(\rho_{c})$. Note that the mass-radius behavior of the
configurations  with surface \textit{crust} resembles those of the quark stars
\cite{Glendening:2000}. In fig. \ref{Jpint} we show the angular
momentum for mass shedding angular velocity as a function of the core-\textit{crust} transition
pressure. In the counter rotating configuration we find that if the core-\textit{crust} transition
pressure is large enough, the angular momentum of the inner \textit{crust}
region could be equal or even higher than the angular momentum of the star
core, resulting in a null or negative total
angular momentum.

\section{Core-\textit{crust} transition pressure evolution in giant
  \textit{glitches} of the Vela pulsar}

On a typical pulsar, the rupture of the solid \textit{crust} produces a release of  
 energy $10^{41}$-$10^{43}$ ergs to the inner part of the \textit{crust}. Several works 
study the thermal response of a neutron star after a \textit{glitch} 
\cite{Hirano:1997,VanRiper:1997}, obtaining that increases in the temperature of the transition region of the
order of  $10^{6} K$. This small rise of the temperature will produce an small
but sensitive increase in the
transition pressure \cite{Prakash:1997rep}. We propose that the \textit{glitch}
   can be explained by changes in the equation of the state of the star in the
   transition region between the \textit{crust}  and the core. In our model,
   these changes are represented by  variations of the transition pressure due
   to the increase of the temperature produced by  the deposition of the
   energy, coming from the rupture of the \textit{crust}.

Using the observational data from the ATNF Pulsar Catalogue
\cite{ATNF_cat} we can study the giant glitches of Vela pulsar using our model.
We assume that during the \textit{glitch} neither the total mass nor the
central density of the star are changed. Hence, in our description,  the
changes in the pulsar during this epoch are due to readjustment of the
core-\textit{crust} transition 
pressure, i.e., a variation of the equation of state in transition region
between the core and the \textit{crust}. Assuming this, we can give a
description of the effects of these \textit{glitches} over the 
neutron star using our model of slowly rotating neutron star with superficial
\textit{crust}. We assume that the star has a total mass of $1.44 M_{\odot}$, a
central density of $1.279\cdot10^{15} gcm^{-3}$, and a transition pressure 
before the \textit{glitch} of $3.751\cdot10^{33} dyn/cm^{2}$. We calculate the
resulting configuration after the \textit{glitch} varying the
core-\textit{crust} transition pressure
until we reach an equivalent configuration (same mass and central density)
with the resulting angular velocity just after the \textit{glitch}. As it can be seen
in table \ref{vela_table}, our model predicts a relative transition pressure
increase of the order 
of $10^{-10}$ for the considered \textit{glitches}.
This small variation in the equation of state on the surface of the star causes a
redistribution of other properties of the star. The
eccentricity of 
the star increases order $10^{-6}$. Also, a movement of matter on the
surface of the star, from the poles to the equator, produce an increase of the
quadrupolar momentum ($10^{-6}$) which may cause gravitational
radiation.       

Using a estimation for the thermal pressure variations
it can be
seen that the corresponding increase of the inner \textit{crust} temperature
is order $10^{6} K$ \cite{Prakash:1997rep}, which agree with the results
obtained in the works previously commented.
So the relative core-\textit{crust} transition pressure
changes we obtain in our 
model can be understood as variations in the
equation of state of the transition region between the core and the
\textit{crust} of the neutron star, resulting from the deposition of
energy of $10^{41}$-$10^{43} erg$ in the outer layers of the neutron star during the \textit{glitch}
time. 


\begin{table*}
\begin{center}
\tabcolsep 2.5pt 
\scriptsize 
\begin{tabular}{c c c c c c }   
Date (MJD) & 
$\delta \Omega (10^{-6})$ & 
$\delta p_{int} (10^{-10})$ & 
$\delta ecc (10^{-6})$ &
$\delta Q (10^{-6})$ \\ [1.0ex]
\hline
40289 & 2.34 & 0.70 & 2.36 & 4.73 \\           
41192 & 2.05 & 0.58 & 1.97 & 3.94 \\
43693 & 3.06 & 0.93 & 3.15 & 6.31 \\
45192 & 2.05 & 0.58 & 1.96 & 3.93 \\
48457 & 2.72 & 0.81 & 2.76 & 5.51 \\
51559 & 3.09 & 0.93 & 3.14 & 6.28 \\
53959 & 2.62 & 0.82 & 2.74 & 5.48 \\ [1.0ex]
\hline                    
\end{tabular}
\end{center}
 \caption{Date of each giant \textit{glitch} of the Vela pulsar in modified
   Julian days and relative increases of angular velocity, transition
   pressure, eccentricity and quadrupolar moment} 
 \label{vela_table}
 \end{table*}



In conclusion, we propose a new model for the giant \textit{glitches} of the Vela
pulsar: The tensions on the \textit{crust} trying to adequate the ellipticity
of the \textit{crust} to the changed angular velocity produce the rupture  
of the \textit{crust}. Then, there is a  sudden energy
deposition in the inner \textit{crust} layers, which causes a rise of its temperature,
resulting in an increase of the core-\textit{crust} transition pressure of the neutron
star (a change in the equation of the state in the core-\textit{crust}
transition region is produced).  Hence, the properties of the neutron star
(angular velocity, eccentricity,  surface matter distribution and quadrupolar
moment) change and a \textit{glitch} is  generated/observed. 

 The present work has been supported by Spanish Ministry of Science 
Project FIS2009-10614. The authors wish to thank F. Navarro-L\'erida for valuable discussions.


\label{lastpage}

\end{document}